\documentstyle{article}
\begin{document}  

\begin{center}
{\bf Decohering quantum statistics: Suppression of BEC}\\
\vspace{0.5cm}

N. KUMAR\\ 
Raman Research Institute, Bangalore 560080, India\\
\end{center}

\begin{abstract}
It is argued that the environment induced decoherence between spatially separated parts of the wavefunction for an open quantum system also classicalizes the quantum statistics which is based on indistinguishability. This may imply a suppression of the BEC upon dilution with, e.g., isotopic fermions, in that the usual phase-space density criterion $n\Lambda^3 >$ 2.612 is supplemented by the condition $n\Lambda^3 > \hbar\gamma/\sqrt{4\pi}k_BT$, $\gamma$ being the friction coefficient subtended by the environment, i.e., the fermions in this case.
\end{abstract}

The idea of decoherence of an open quantum system of interest due to its interaction, or entanglement with the unobserved, but again quantum-mechanical, degrees of freedom of the environment, has informed much of the on-going discussion on the deep problem of emergence of a classical behaviour out of what is fundamentally quantum mechanical in nature [1-3]. Thus, e.g., we have the well posed question as to when, if at all, can we have a superposition of the two alternative, macroscopically distinct, quantum states. The mechanism of decoherence resolves this issue within the framework of unitary evolution in quantum mechanics, without invoking a wavefunction collapse [3-4]. Here the environmental degrees of freedom effectively {\it monitor} some of the system observables, whose eigenstates then decohere continuously. (In terms of the reduced density matrix of the subsystem of interest, decoherence implies damping/nulling of its off-diagonal matrix elements, while the non-vanishing diagonals represent the preferred (pointer) basis determined by the detector). For a macroscopic open system, with necessarily large number of internal degrees of freedom, the decoherence time is generally too short to access experimentally (see reference [2] for a lucid discussion on this point). For large molecules e.g., the fullerenes, highly sophisticated experiments have recently  been carried out that clearly demonstrate controlled decoherence [5].  The decoherence time-scales are, however, experimentally readily accessible for the phase-sensitive studies on microscopic and mesoscopic systems, as in weak localization in disordered conductors [6] and flux tunneling in Josephson functions [4]. Here the decoherence is often referred to\,\, as\,\, dephasing, \,\,or \,\,phase \,\,breaking. \,\,\,A \,\,\,striking\,\, example\,\, of\,\, decoherence
in condensed matter that has recently been treated theoretically is the Landau orbital diamagnetism, a purely quantum phenomenon, that was shown to be suppressed by decoherence caused by a dissipative coupling to the environment [7]. In the present work, we have addressed yet another aspect of decoherence $-$ namely, the decoherence of {\it quantum statistics}, in particular of Bose statistics, which is conceptually distinct from, though physically related to, the usual idea of wavefunction decoherence referred to above. Our main result here is that the decoherence of quantum statistics can lead to depression/suppression of BEC [8] in a nearly ideal Bose gas upon dilution with, e.g., the isotopic fermions, that can provide the environmental monitoring of the Bose system of interest. 

Let us first introduce the conceptual basis of our argument. In quantum mechanics, we hold that the classically identical particles are quantum-mechanically indistinguishable. Following Landau and Lifshitz [9]. This is best appreciated by viewing the indistinguishability as a consequence of our inability, in principle, to tag and track the classically identical particles inasmuch as quantum mechanics does not admit particle trajectories. This is where decoherence comes in. Under  persistent {\it monitoring} by the environment, the particle wavefunction decoheres continually to give nearly classical trajectories, making the identical particles operationally distinguishable by their trajectories, now become identifiable. Hence the classicalization of quantum statistics. The argument is admittedly heuristic, but has a compelling logic to it, and motivates further enquiry. 

In order to treat the idea of decoherence of statistics further, it is apt to first consider a minimal system of two identical (Bose) particles, and introduce the Wigner distribution $W(\Delta {\bf x}, \Delta {\bf p})$ corresponding to their {\it relative} wavefunction. More specifically, consider the minimum-uncertainty Wigner function
\begin{equation}
W(\Delta {\bf x}, \Delta {\bf p}) = \left(\frac{1}{\pi\hbar}\right )^3 exp\left (- \frac{(\Delta {\bf x} - \Delta {\bf x}_0)^2}{\delta^2} - \frac{(\Delta {\bf p} - \Delta {\bf p}_0)^2}{\hbar^2} \delta^2\right ),
\end{equation}
with $\delta$ the minimum-uncertainty parameter for the relative wavepacket. Here $\Delta {\bf x}_0$, $\Delta {\bf p}_0$ are, respectively, the centres of the relative position, momentum, with $\Delta x_0 \gg \delta, \Delta p_0 \gg \hbar/\delta$. Now, for the ideal Bose gas of number density $n$, we can identify $\Delta x_0$ with the mean inter-particle spacing $\sim n^{-1/3}$, and $\Delta p_0$ with the typical momentum associated with the thermal de Broglie wavelength $\Lambda_{th}$, i.e.,
\begin{eqnarray}
\Delta p_0 &\sim& 2\pi\hbar/\Lambda_{th}, \nonumber\\
\Lambda_{th} &=& (2\pi\hbar^2/2m k_BT)^{1/2}.
\end{eqnarray}
Quantum statistics would then demand that the time taken to traverse the relative separation $\sim \Delta x_0$ at the relative velocity $\Delta p_0/m$ should be less than the wavefunction decoherence time $\tau_D$ due to environmental interaction. Thus, we have the condition for the quantum statistics to hold effectively, as 
\begin{equation} 
\left (\frac{\Delta p_0}{m}\right ) \tau_D > \Delta x_0.
\end{equation}
Equations (2) and (3), together with $\Delta x_0 \sim n^{-1/3}$ give the condition for the coherence of quantum statistics for a 3D system of Bose gas of number density $n$ and temperature $T$ in terms of the wavefunction decoherence time $\tau_D$. The latter can now be estimated from a model of decoherence. Thus, e.g., consider the case of a gas of particles whose x-coordinates are coupled to a scalar field $\phi(q,t)$ propagating orthogonal to $x$, via a potential $\epsilon x d\phi/dt$ [10].  It is known for this case, that the decoherence time $\tau_D$, for the parts of the wavefunction separated spatially by a distance $\Delta x$, is given by [10] 
\begin{equation}
\tau_D = (1/4\pi\gamma) \left(\frac{\Lambda_{th}}{\Delta x}\right )^2
\end{equation}
with $\gamma = \eta/2m$, where $\eta = \epsilon^2/2$ is the viscosity offered by coupling to the environment. (The factor of 4$\pi$ is due to  a slightly different definition of $\Lambda_{th}$ in ref. 10). Equation (4) is expected to hold quite generally as far as the time- and the length-scale estimates are concerned. For our purpose now, the environment can be realized, e.g., by diluting the Bose gas with the isotopic fermions. The length scale $\Delta x$ in our case can then be identified with the mean inter particle spacing, i.e., $\Delta x \sim n^{-1/3}$.  From Eqs. (3) and (4), the condition for the quantum (Bose) statistics to hold effectively, then turns out to be
\begin{equation}
n\Lambda_{th}^3 > \frac{\sqrt{4\pi}\hbar \gamma}{k_BT} .
\end{equation}
For the BEC to occur, the condition expressed by Eq. (5) must be satisfied simultaneously with the well-known phase-space density criterion $n\Lambda_{th}^3 \geq$ 2.612. Thus, for a frictional coefficient $\gamma > 2.614 \times k_BT/\sqrt{4\pi}\hbar \sim 100 s^{-1}$ at $T\sim 1$ nK, we expect an appreciable suppression of BEC.  It is, however, not clear to us at the moment if a frictional parameter of this order can be readily realized experimentally by isotopic dilution. But, it is well within the reach of some form of optical molass-ing. A near-ideal BEC in a dilute alkali gas, with the condensation fraction $\approx$ 1, seems to be the best experimental system in which to look for this effect of decoherence of statistics. It is important to note here that the decoherence of quantum statistics by the environmental coupling (represented by the friction $\gamma >$ 0) is {\it not} the same as the statistical thermal effect $-$ the high-temperature limit of any quantum statistics is, of course, classical. This statistical population effect is, however,distinct from the dynamical decoherence for $\gamma >$ 0 as discussed here. 

In conclusion, to the best of our knowledge, we have for the first time extended the idea of decoherence of particle wavefunction to that of the decoherence of quantum statistics. We have considered this idea in the context of the most striking manifestation of quantum statistics, namely the BEC. A nearly ideal BEC in an alkali gas diluted with isotopic fermions is expected to be a good experimental system to show this decoherence as a  depression/suppression of BEC.

\end{document}